\documentclass[slac_one]{revtex4}
\usepackage{graphicx}
\usepackage{fancyhdr}
\begin{document}

\title{Measurements of multijet production at low-{\boldmath $x$}} 

%

\author{J. Terron (on behalf of the H1 and ZEUS Collaborations)}
\affiliation{Departamento de F\'{\i}sica Te\'orica C-XI, Facultad de Ciencias, Universidad Aut\'onoma de Madrid, Cantoblanco, Madrid 28049, Spain}

\begin{abstract}
Recent measurements of multijet production in neutral current deep
inelastic $ep$ scattering at HERA are presented. Emphasis is put 
on parton dynamics at low $x$.
\end{abstract}

\maketitle

\thispagestyle{fancy}


\section{INTRODUCTION}
Deep inelastic scattering (DIS) off protons has provided decisive information on
the parton distribution functions (PDFs) of the proton. Inclusive
measurements of the cross section for the reaction $ep\rightarrow e{\rm X}$
as a function of the virtuality of the exchanged boson ($Q^2$) and of the
Bjorken scaling variable ($x$) have been used to determine the proton
structure function $F^p_2(x,Q^2)$. Perturbative QCD (pQCD) in the next-to-leading-order
(NLO) approximation has been widely used to extract the proton PDFs from such
measurements and to test the validity of pQCD.
In the standard approach (DGLAP~\cite{dglap}), the evolution equations sum up
all leading double logarithms in $\ln{Q^2}\cdot \ln{1/x}$ along with single
logarithms in $\ln{Q^2}$ and are expected to be valid for $x$ not too small.
At low $x$, a better approximation is expected to be provided by the BFKL
formalism~\cite{bfkl} in which the evolution equations sum up all leading double
logarithms along with single logarithms in $\ln{1/x}$.

The DGLAP evolution equations have been tested extensively at HERA and were
found to describe, in general, the data. In particular, the striking rise of the measured
$F^p_2(x,Q^2)$ at HERA with decreasing $x$ can be accomodated in the DGLAP
approach. Nevertheless, the inclusive character of $F^p_2(x,Q^2)$ may obscure 
the underlying dynamics at low $x$, and more exclusive final states
like forward~\footnote{The coordinate system used is a right-handed
Cartesian system with the $Z$ axis pointing in the proton beam direction, referred
to as the ``forward direction''.} jets~\cite{mueller} need to be studied.
BFKL evolution predicts a larger fraction of small-$x$ events containing high-$E_T$ forward 
jets than predicted by DGLAP~\cite{mueller,otros}. Parton dynamics at low $x$ 
is particularly relevant for the LHC given that most of the interesting hard
processes involve partons with low fractional momenta. 

\section{FORWARD JET PRODUCTION}
\begin{figure*}[h]
\setlength{\unitlength}{1.0cm}
\begin{picture} (10.0,8.0)
\put (-5.0,0.0){\includegraphics[width=70mm]{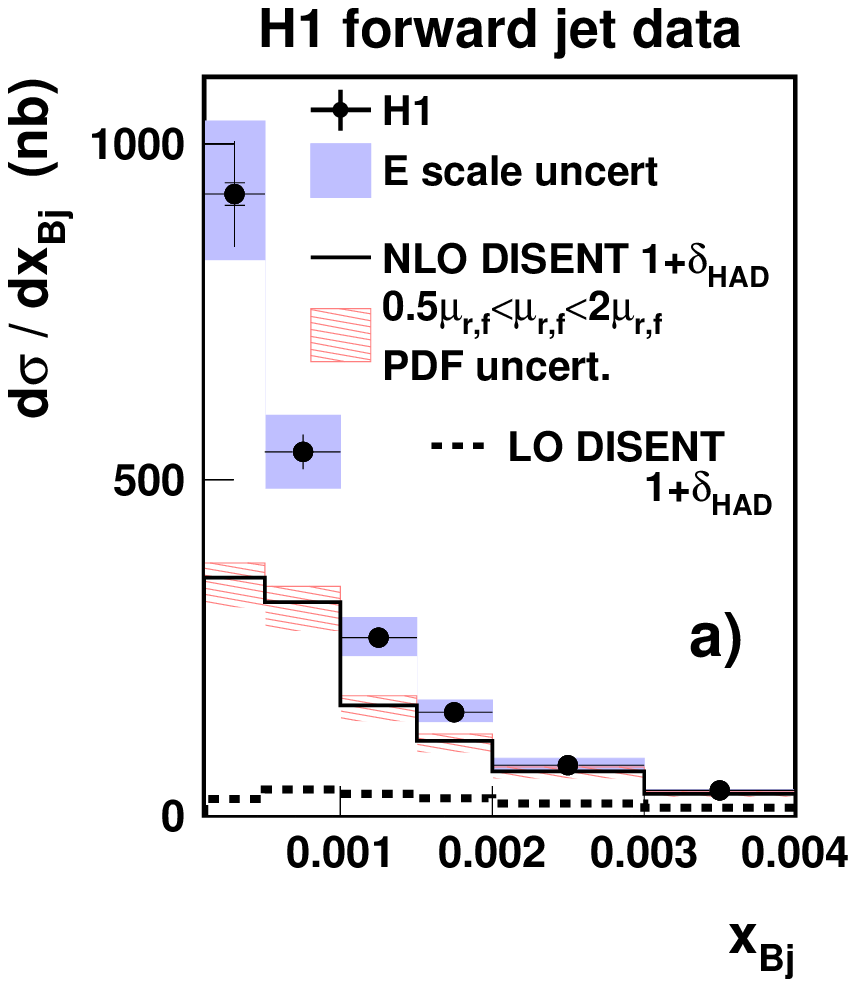}}
\put (2.0,2.0){\includegraphics[width=50mm]{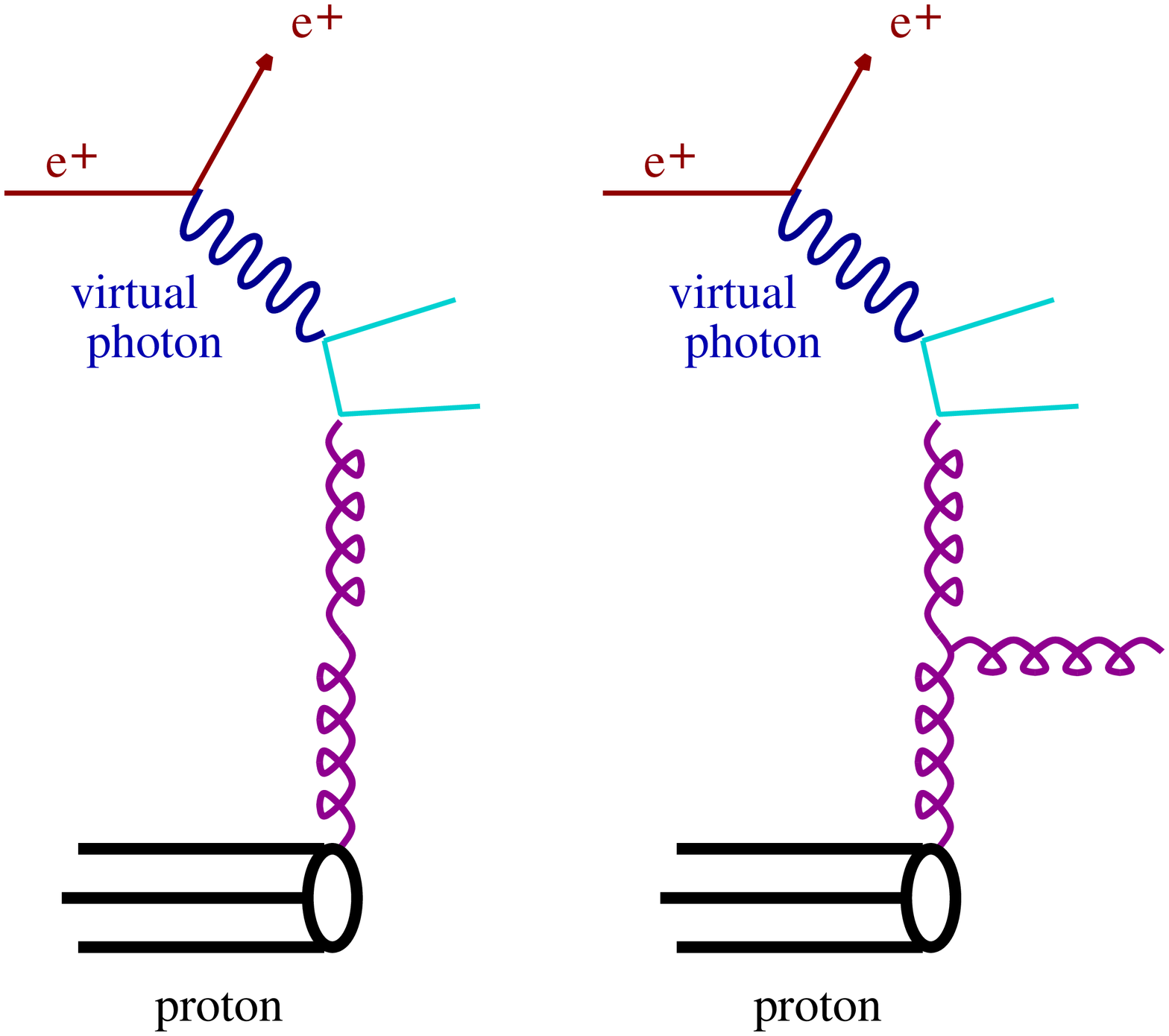}}
\put (7.1,-0.3){\includegraphics[width=60mm]{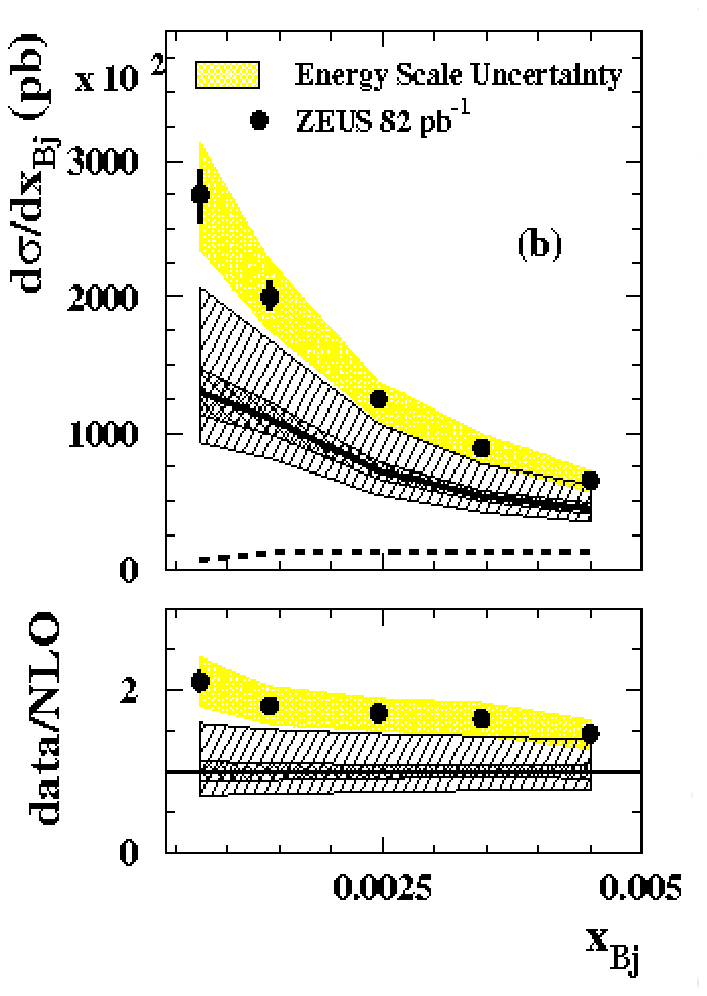}}
\put (-1.1,-0.3){(a)}
\put ( 4.1,-0.3){(b)}
\put (10.1,-0.3){(c)}
\put (2.3,1.3){LO ${\cal O}(\alpha_s)$}
\put (5.0,1.3){NLO ${\cal O}(\alpha^2_s)$}
\end{picture}
\caption{Measurements of forward jet production in NC DIS as functions of
$x$ (a,c). Examples of Feynman diagrams (b).} \label{figura1}
\end{figure*}

Forward jet production in neutral current (NC) DIS has been studied
extensively at HERA. As an example, measurements of forward jet production with 
$p_{t,{\rm jet}}> 3.5$~GeV, polar angle $\theta_{\rm jet}$ between $7^{\circ}$ 
and $20^{\circ}$, $0.5 < p^2_{t,{\rm jet}}/Q^2 < 5$ and 
$x_{\rm jet}\equiv E_{\rm jet}/E_p>0.035$, where $E_p$ is the proton-beam energy, 
in the kinematic region defined by $10^{-4} < x < 4 \cdot 10^{-3}$ and 
$5 < Q^2 < 85$~GeV$^2$ are shown in Figure~\ref{figura1}a and exhibit a strong 
rise towards low $x$~\cite{h1forw}. Perturbative QCD apparently does not account 
for such a rise: the leading-order (LO) QCD calculation does not predict any rise 
while the one predicted by NLO~\cite{disent} is still much too low (see 
Figure~\ref{figura1}a). 
Some of the Feynman diagrams that are accounted for in the LO 
(${\cal O}(\alpha_s)$) and NLO (${\cal O}(\alpha_s^2)$) calculations are shown in 
Figure~\ref{figura1}b. The former has no additional gluon radiation and helps to 
understand why in the LO calculation there is hardly any phase space left for 
forward jet production. In contrast, the NLO calculation accounts for the
radiation of one additional gluon. This explains why there is such a huge increase from 
LO to NLO: due to the opening of a new channel, namely gluon-exchange in the $t$ channel. 
However, that means that the NLO calculation is effectively a ``LO'' calculation, since 
no corrections are included. The NLO calculation should thus have large theoretical 
uncertainties from higher orders. A variation of the renormalisation scale around 
$\mu^2_R=\langle p^2_{t,{\rm dijets}}\rangle$ do not give rise to such large theoretical
uncertainties (see Figure~\ref{figura1}a). However, if $Q^2$ is instead chosen as 
the renormalisation scale, the resulting theoretical uncertainties are large, as pointed
out in~\cite{h1forw} and shown in Figure~\ref{figura1}c:
measurements of forward jet production~\cite{zeusforw} with 
$E^{\rm jet}_T>5$~GeV, pseudorapidity in the range $2 < \eta^{\rm jet}<4.3$,
$0.5 < (E^{\rm jet}_T)^2/Q^2<2$ and $x_{\rm jet}>0.036$ in the kinematic region
given by $4 \cdot 10^{-4} < x < 5 \cdot 10^{-3}$ and $20 < Q^2 < 100$~GeV$^2$ are
compared to NLO QCD calculations~\cite{disent} with $\mu^2_R=Q^2$. Large theoretical
uncertainties which arise from higher orders in the pQCD calculations prevent a firm
conclusion. Further progress can be made by making measurements for which genuine NLO
calculations are available, i.e. one gluon radiation at LO and two additional radiated
gluons at NLO. That is the case for three-jet production, which was already studied
in~\cite{h1forw} and has been investigated more thoroughly in~\cite{h1three}. The latter 
is discussed next.

\section{MULTIJET PRODUCTION AT LOW {\boldmath $x$}}
\begin{figure*}[h]
\setlength{\unitlength}{1.0cm}
\begin{picture} (10.0,7.0)
\put (-4.1,1.0){\includegraphics[width=70mm]{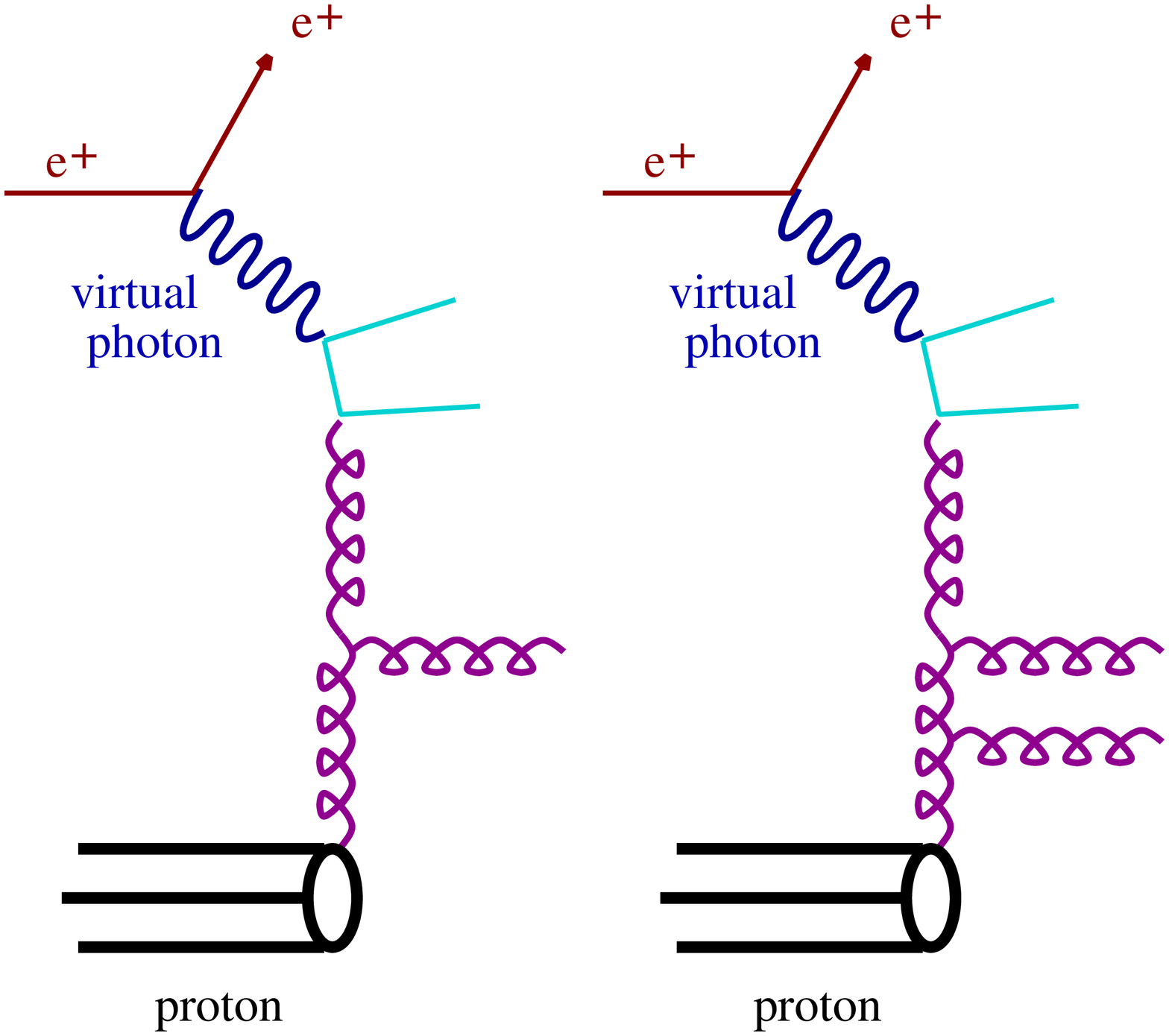}}
\put (3.1,0.7){\includegraphics[width=60mm]{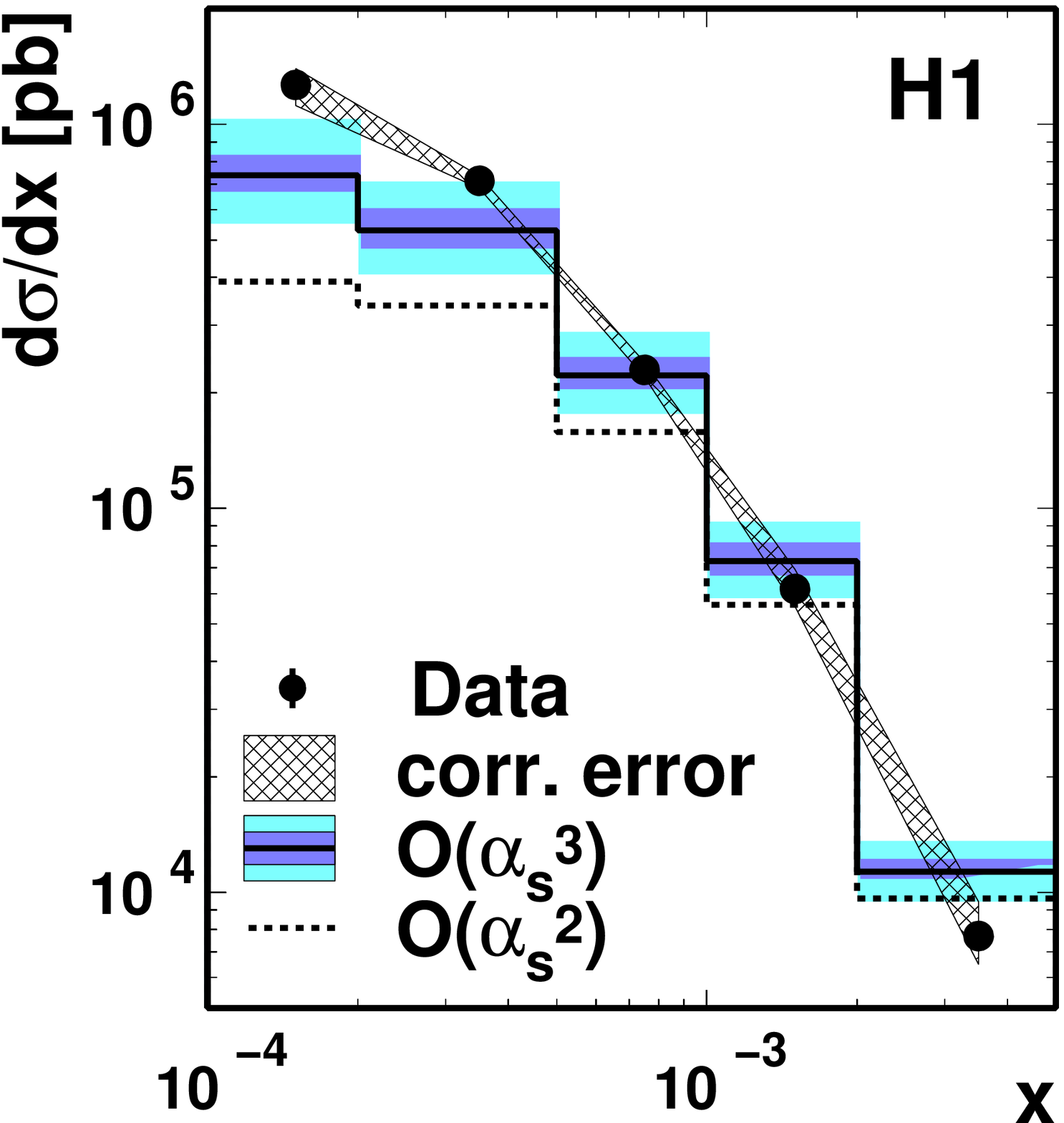}}
\put (10.1,2.3){\includegraphics[width=40mm]{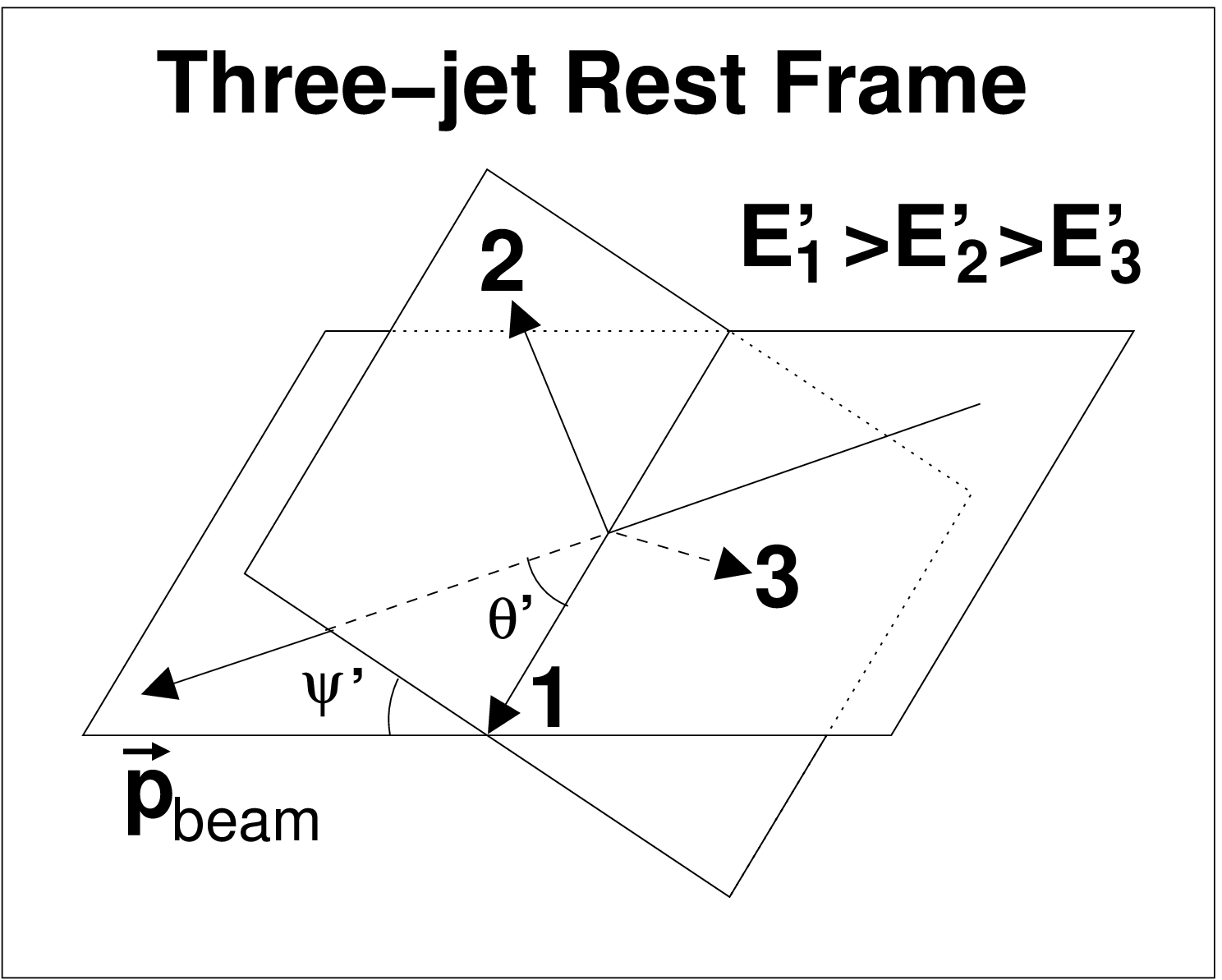}}
\put (-1.1,-0.3){(a)}
\put ( 6.1,-0.3){(b)}
\put (12.1,-0.3){(c)}
\put (-3.3,0.3){LO ${\cal O}(\alpha^2_s)$}
\put (0.0,0.3){NLO ${\cal O}(\alpha^3_s)$}
\end{picture}
\caption{Examples of Feynman diagrams (a). Measurement of $d\sigma/dx$ for 
three-jet production in NC DIS as a function of $x$ (b). 
Definition of $\theta^{\prime}$ and $\psi^{\prime}$ in the three-jet 
centre-of-mass frame (c).} \label{figura2}
\end{figure*}

Some of the Feynman diagrams accounted for in the pQCD calculations for three-jet 
production in NC DIS are shown in Figure~\ref{figura2}a. A measurement of the
differential cross section $d\sigma/dx$ as a function of $x$ for three-jet
production~\cite{h1three} is presented in Figure~\ref{figura2}b. The jets are 
reconstructed in the $\gamma^{*}p$ frame and are required to fulfill the following 
conditions: the transverse momentum of each jet $p_{t,i} >4$~GeV, the sum of the
highest and next-to-highest $p_t$ jets above 9~GeV, the jet pseudorapidity in the
laboratory frame to lie between -1 and 2.5 and at least one of the jets in the central 
region ($-1 < \eta^{\rm lab}_{\rm jet}<1.3$). The kinematic region is defined
by $10^{-4} < x < 10^{-2}$, $5 < Q^2 < 80$~GeV$^2$ and $0.1 < y <0.7$, where $y$ is
the inelasticity variable. Perturbative QCD calculations are compared to the data
in Figure~\ref{figura2}b: the LO (${\cal O}(\alpha_s^2)$) calculation still falls
short of the data, but the NLO (${\cal O}(\alpha_s^3)$) calculation~\cite{nlojet}
improves dramatically the description of the data at low $x$.

\begin{figure*}[h]
\setlength{\unitlength}{1.0cm}
\begin{picture} (10.0,4.5)
\put (-4.1,-0.3){\includegraphics[width=40mm]{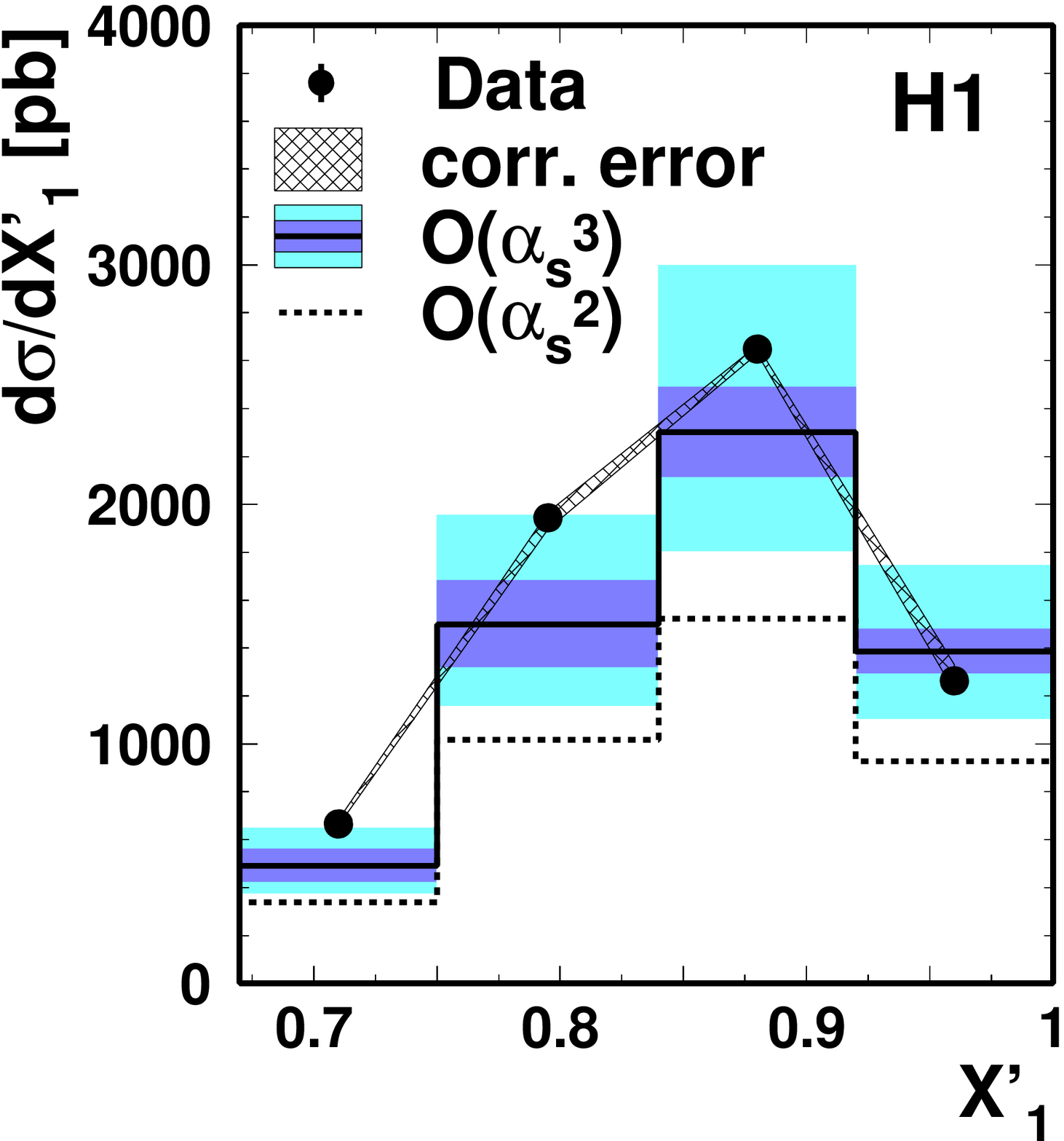}}
\put (0.6,-0.3){\includegraphics[width=40mm]{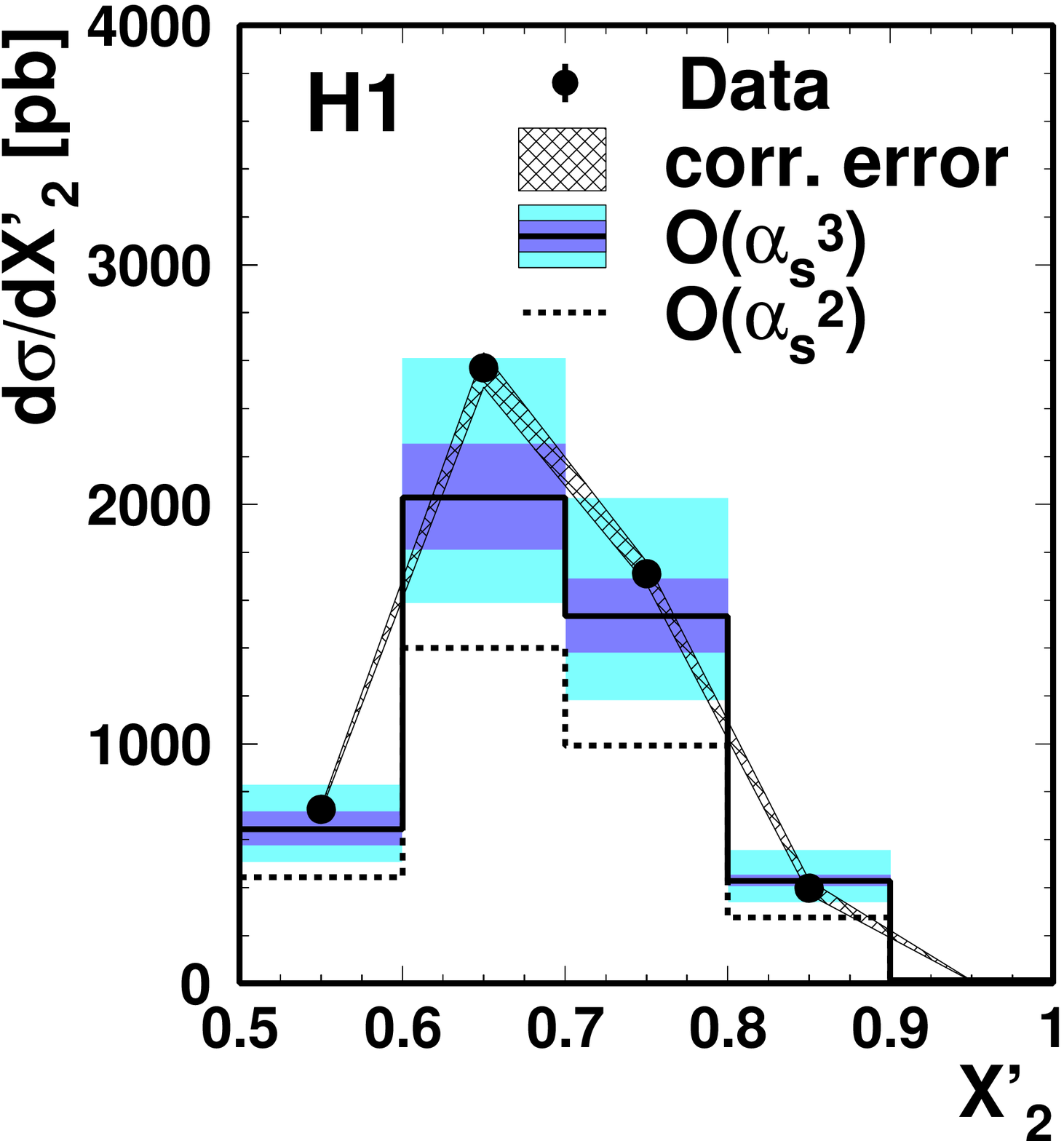}}
\put (5.1,-0.3){\includegraphics[width=40mm]{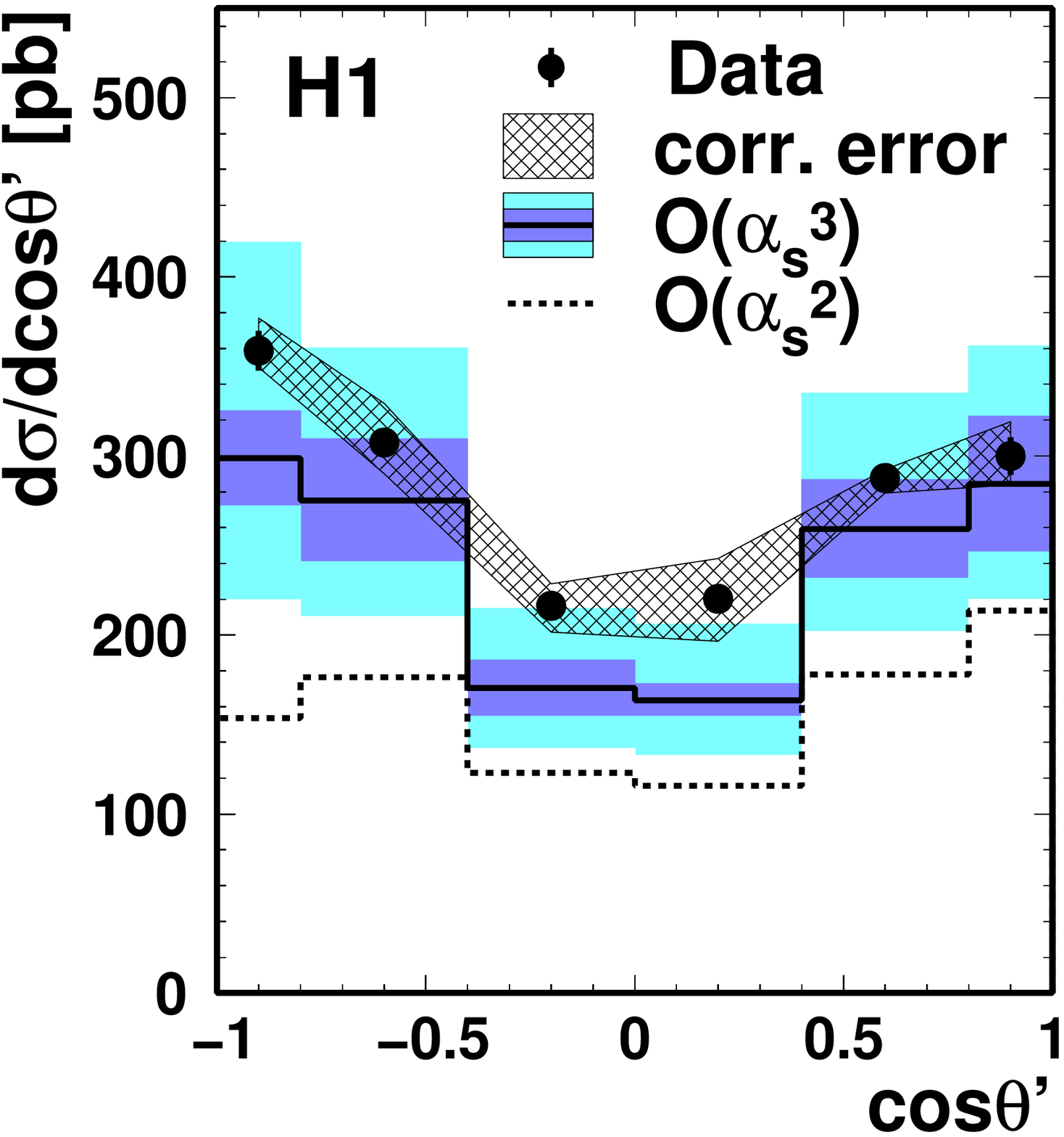}}
\put (9.8,-0.3){\includegraphics[width=40mm]{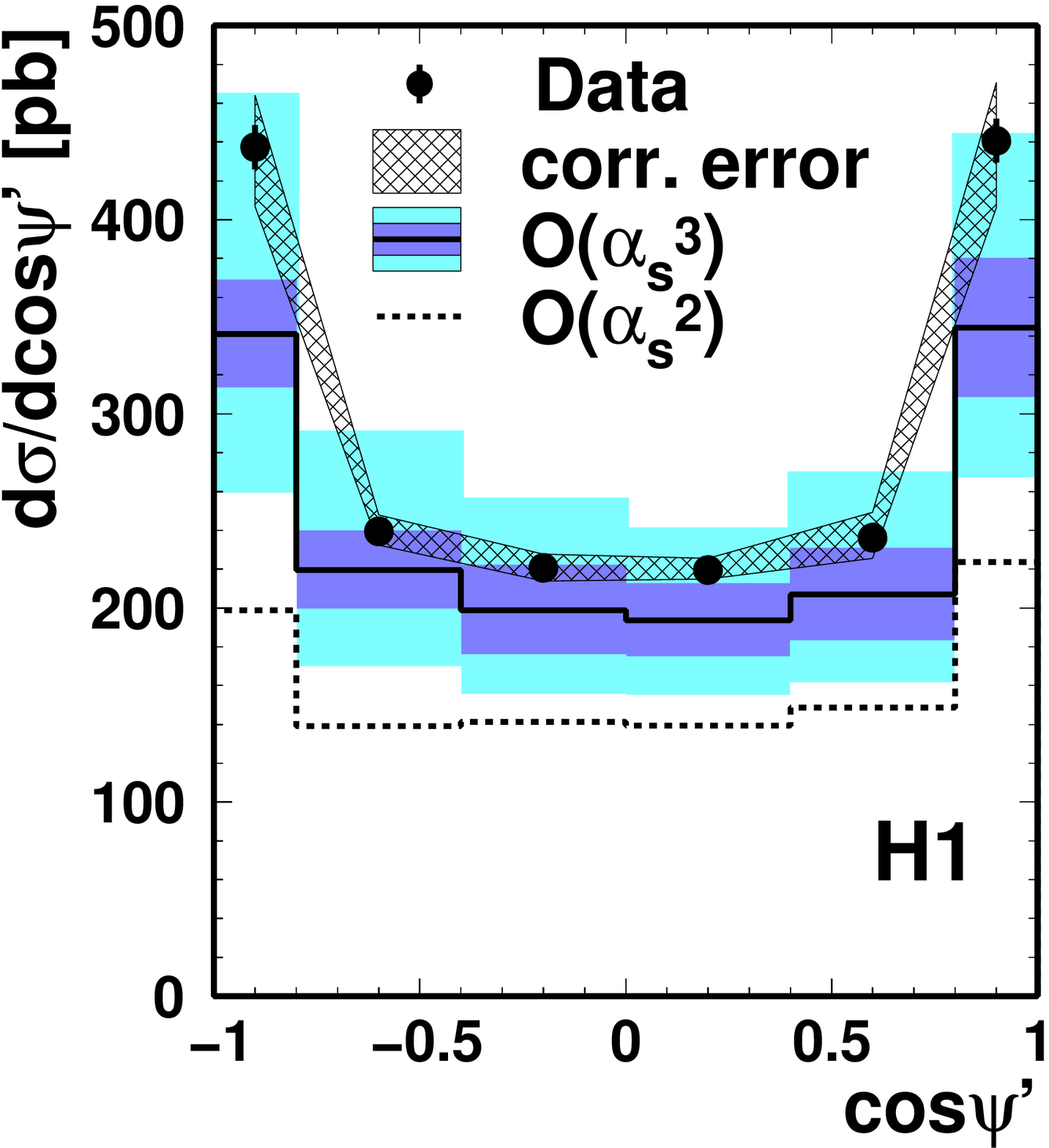}}
\end{picture}
\caption{Measurements of the differential cross sections as functions of the 
variables used to describe the topology of three-jet events in the three-jet 
centre-of-mass frame.} \label{figura3}
\end{figure*}

The inclusion of the ${\cal O}(\alpha_s^3)$ QCD corrections does not only improve 
the description of the measured rate but also that of the topology of the events. 
Measurements have been made of the distributions in the variables used to describe the
topology of three-jet events in the three-jet centre-of-mass frame: the scaled
energy of the jets 
$X^{\prime}_i\equiv 2E^{\prime}_i/(E^{\prime}_1+E^{\prime}_2+E^{\prime}_3)$
($i=1,2$; $E^{\prime}_1 > E^{\prime}_2 > E^{\prime}_3$) and the two angles
$\theta^{\prime}$ and $\psi^{\prime}$ (see Figure~\ref{figura2}c). The measurements
are shown in Figure~\ref{figura3} and compared to NLO calculations~\cite{nlojet}. The 
inclusion of additional gluon radiation provides an improved description of the data,
for example, in the $\cos{\theta^{\prime}}$ distribution: the NLO follows the data 
and exhibits peaks at $\cos{\theta^{\prime}}=-1$ and $1$, whereas the LO calculation 
flattens out at $\cos{\theta^{\prime}}=-1$.

\begin{figure*}[t]
\setlength{\unitlength}{1.0cm}
\begin{picture} (10.0,12.50)
\centering
\put (0.0,-0.3){\includegraphics[width=100mm]{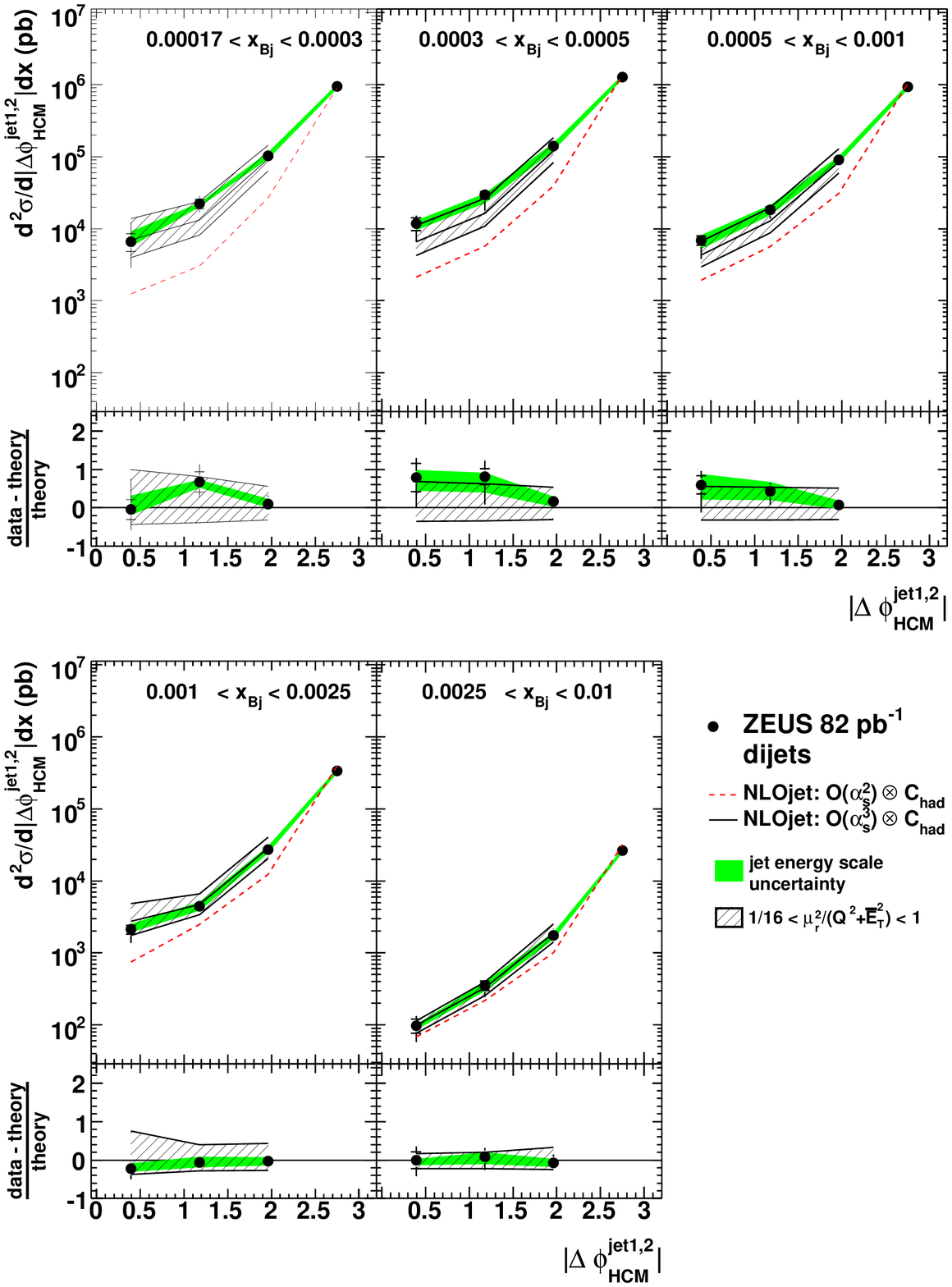}}
\end{picture}
\caption{\mbox{Measurements of the doubly differential cross section 
$d^2d\sigma/d|\Delta\phi^{jet 1,2}_{\rm HCM}|dx$ for dijet production 
in different regions of~$x$.}} \label{figura4}
\end{figure*}

Further investigations of low-$x$ parton dynamics have been made by
studying transverse-energy and angular correlations in dijet and trijet
production in NC DIS~\cite{zeusthree}. Jets are reconstructed in
the hadronic centre-of-mass (HCM) frame and required to fulfill the
following conditions: $E^{\rm jet 1}_{T,{\rm HCM}}>7$~GeV,
$E^{\rm jet 2,3}_{T,{\rm HCM}}>5$~GeV and $-1 < \eta^{jet 1,2,3}_{\rm lab}<2.5$.
The kinematic region is given by $10^{-4} < x < 10^{-2}$, 
$10 < Q^2 < 100$~GeV$^2$ and $0.1 < y <0.6$. One of the most interesting
angular correlations is provided by the variable $|\Delta\phi^{jet 1,2}_{\rm HCM}|$,
which is defined as the azimuthal separation of the two jets with largest
$E^{\rm jet}_{T,{\rm HCM}}$. For dijet events, ${\cal O}(\alpha_s)$ kinematics
constrain $|\Delta\phi^{jet 1,2}_{\rm HCM}|$ to $\pi$ and ${\cal O}(\alpha^2_s)$
calculations provide the LO contribution; ${\cal O}(\alpha^3_s)$
calculations give the NLO correction.
Measurements of the doubly differential cross section 
$d^2d\sigma/d|\Delta\phi^{jet 1,2}_{\rm HCM}|dx$ for dijet production 
in different regions of $x$ are presented in Figure~\ref{figura4}.
The ${\cal O}(\alpha^2_s)$ predictions increasingly deviate from the data
as $x$ decreases, whereas ${\cal O}(\alpha^3_s)$ calculations~\cite{nlojet} provide a
good description of the data even at low $x$.

In summary, parton dynamics at low $x$ is vigorously pursued at HERA. Precise
measurements of multijet production in NC DIS have been made down to $x\sim 10^{-4}$ 
in terms of jet rates, topologies and correlations. Comparison of perturbative
QCD calculations with these measurements demonstrate the big impact of initial-state
gluon radiation. Perturbative QCD at ${\cal O}(\alpha_s^3)$ reproduces succesfully
the measurements. However, the theoretical uncertainties are still significant
and the precision of the data demands next-to-next-to-leading-order corrections to 
be included.

\begin{thebibliography}{99}   
\bibitem{dglap}
V.N.~Gribov and L.N.~Lipatov, Sov. J. Nucl. Phys. 15 (1972) 438;\\
L.N.~Lipatov, Sov. J. Nucl. Phys. 20 (1975) 94;\\
Yu.L.~Dokshitzer, Sov. Phys. JETP 46 (1977) 641;\\
G. Altarelli and G. Parisi, Nucl. Phys. B126 (1977) 298.
\bibitem{bfkl}
E.A.~Kuraev, L.N.~Lipatov and V.S.~Fadin, Sov. Phys. JETP 45 (1977) 199;\\
Ya.Ya.~Balitski\u i and L.N.~Lipatov, Sov. J. Nucl. Phys. 28 (1978) 822.
\bibitem{mueller}
A.H.~Mueller, Nucl. Phys. Proc. Suppl. C18 (1991) 125;\\
A.H.~Mueller, J. Phys. G17 (1991) 1443.
\bibitem{otros}
J.~Bartels, A.~De Roeck and M.~Loewe, Z. Phys. C54 (1992) 645;\\
J.~Kwiecinski, A.D.~Martin and P.J.~Sutton, Phys. Lett. B287 (1992) 254;\\
W.K.~Tang, Phys. Lett. B278 (1992) 363.
\bibitem{h1forw}
H1 Collaboration, A.~Aktas et al., Eur. Phys. J. C46 (2006) 27.
\bibitem{disent}
S.~Catani and M.H.~Seymour, Nucl. Phys. B485 (1997) 291.
\bibitem{zeusforw}
ZEUS Collaboration, S.~Chekanov et al., Eur. Phys. J. C52 (2007) 515.
\bibitem{h1three}
H1 Collaboration, F.D.~Aaron et al., Eur. Phys. J. C54 (2008) 389. 
\bibitem{nlojet}
Z.~Nagy and Z.~Trocsanyi, Phys. Rev. Lett. 87 (2001) 082001.
\bibitem{zeusthree}
ZEUS Collaboration, S.~Chekanov et al., Nucl. Phys. B786 (2007) 152.
\end{thebibliography}
\end{document}